\PassOptionsToPackage{square,numbers}{natbib} 
\documentclass[10pt]{article}

\usepackage[utf8]{inputenc}
\usepackage[T1]{fontenc}
\usepackage[english]{babel}
\usepackage{graphicx}
\usepackage{amsmath,amssymb,amsbsy,amsthm,color,natbib}
\usepackage{multicol}
\usepackage{color} 
\usepackage{natbib}
\usepackage[linktoc=all,colorlinks=true,citecolor=blue,linkcolor=blue]{hyperref}
\usepackage[paperwidth=18cm,paperheight=29.7cm,hmargin=2cm,vmargin=2.5cm]{geometry}
\usepackage{authblk}
\usepackage{comment}
\usepackage{appendix}
\usepackage{bm}
\usepackage{booktabs}
\usepackage{tabularx}

\graphicspath{{Figures/}}

\title{Chaoticus: a parallel approach to the computation of chaos indicators}

\author[1]{Javier Jiménez-López\thanks{javier.jimenezl@edu.uah.es (Corresponding author)}}
\author[2]{José Sáez-Landete\thanks{jose.saez@uah.es}}
\author[1]{V. J. García-Garrido\thanks{vjose.garcia@uah.es}}
\affil[1]{Departamento de F\'isica y Matem\'aticas, Facultad de Ciencias, Universidad de Alcal\'a, 28805 Alcal\'a de Henares, Madrid, Spain.}
\affil[2]{Signal Theory and Communications Department, University of Alcal\'a, Madrid, Spain.}

\begin{document}

\maketitle


\begin{abstract}

In this paper we present Chaoticus, a Python-based package for the GPU-accelerated integration of ODE systems and the computation of chaos indicators, including SALI, GALI, Lagrangian Descriptors based indicators and the Lyapunov exponent spectrum. By leveraging GPU parallelization, our package significantly reduces the computation times by several orders of magnitude compared to CPU-based approaches. This significant reduction in computing time facilitates the generation of extensive datasets, crucial for the in-depth analysis of complex dynamics in Hamiltonian systems.

\end{abstract}

\noindent \textbf{keywords:} Chaos indicators, Hamiltonian dynamics, Lagrangian descriptors, Parallel computing.

\section{Summary} \label{sec:sum}

The emergence of chaos in deterministic dynamical systems has a profound implication across multiple scientific and engineering disciplines. From the unpredictability of weather patterns \cite{tsonis1989} and the motion of planets \cite{shevchenko2020, dvorak2005} to chemical reactions \cite{gaspard1999}, understanding and detecting chaos has become crucial for both theoretical advancements and practical applications. Hamiltonian systems, which constitute a broad set of conservative physical systems, are very rich in their dynamical behaviour. The phase space of this class of systems exhibits a mixture of chaotic and regular trajectories, and it is the nature of this structure that governs the system's overall behaviour and predictability.

Investigating the structures that are present in phase space and identifying chaotic regions within Hamiltonian systems relies on the numerical integration of the trajectories and the computation of the so called chaos indicators. Several methodologies and software tools have been developed to tackle these problems, as \cite{Aguilar-Sanjuan2021}, where a Python library was presented to reveal phase space structures in Hamiltonian systems using Lagrangian Descriptors \cite{mancho2013,lopesino2017}. However, the computational demands associated with the accurate simulation of Hamiltonian systems, especially when exploring large ensembles of initial conditions or high-dimensional systems, remains significantly challenging.

To address this, we have developed Chaoticus, a Python package created to reduce the computation time required to analyse continuous Hamiltonian systems by using GPU-accelerated functions for the integration of trajectories in phase space while providing robustness when it comes to energy conservation.

\section{Statement of need} \label{sec:son}

Chaos indicators are mathematical tools that allow the determination of a trajectory's nature. Usually, they require information from the trajectory's evolution itself or of the deviation vectors, which satisfy the variational equations of the system \cite{Skokos2010b}.

In the first class of indicators we find Laskar's frequency map analysis \cite{Laskar1993} or the Bikhoff averages method \cite{Levanjic2010,Levanjic2015,Das2016} among others. Then, there are those indicators that require the analysis of the deviation vector's evolution. The following indicators lie in this category: the fast Lyapounov indicator \cite{Froeschle1997a,Froeschle1997b}, the maximal Lyapunov characteristic exponent \cite{Skokos2010a}, the Smaller Alignment Index (SALI) \cite{Skokos2001,Skokos2003,Skokos2004}, the Generalize Alignment Index (GALI) \cite{Skokos2008,moges2025}, which is a generalization of SALI, and the Mean Exponential Growth of Nearby Orbits (MEGNO) \cite{Cincotta2000}. For a comprehensive review of these methods, the reader is referred to \cite{skokos2016chaos}.

All of these chaos indicators are computationally demanding, and those that require the evolution of the deviation vectors have the additional complication of obtaining the variational equations, which has been shown to be a non trivial task \cite{Hillebrand2019}. In spite of their drawbacks, we have decided to implement some of the chaos indicators based on the analysis of the deviation vector's evolution as they are considered to be ground truth indicators, meaning that if the integration time is enough, they will provided the true nature of the trajectories under analysis.

To tackle these issues, chaos indicators based on the framework of Lagrangian Descriptors have been recently introduced \cite{daquin22,Hille22,zimper23,caliman2025,jiménezlópez2025}. This methodology has been proven to be efficient in the task of chaos detection while also being straight forward to implement as there is no need to calculate the variational equations of the system but rather the Lagrangian Descriptor must be calculated alongside the trajectory's evolution \cite{mancho2013,lopesino2017}. This advantage added to the computation time reduction gained with GPU-acceleration gives the capability of conducting detailed analysis of Hamiltonian systems considerably faster than before, therefore, improving the ability of researchers to understand physical phenomena in this wide class of physical systems.

\section{Functionalities} \label{sec:func}

As we have already stated in the previous sections, the aim of this Python-based package is to integrate systems of ODEs that are derived from a Hamiltonian function and compute various chaos indicators that allow the classification of the simulated trajectories using GPU acceleration. To do so, various kernels have been written to integrate the ODE systems using the $8^{th}$ order ODE solver given in \cite{prince1981}. Additionally, the implementation of chaos indicators and the necessary auxiliary functions have also been implemented. 

The package incorporates the following functionalities:
\begin{itemize}
    \item DOP $8$ ODE solver with fixed step size,
    \item DOP $8(5)$ ODE solver with adaptive step size,
    \item DOP $8$ ODE solver with fixed step size that includes the normalization of the deviation vectors,
    \item DOP $8(5)$ ODE solver with adaptive step size that includes the normalization of the deviation vectors,
    \item DOP $8$ ODE solver with fixed step size that includes the normalization of the deviation vectors and allows the computation of the Lyapunov exponents spectrum via QR factorization, 
    \item DOP $8(5)$ ODE solver with adaptive step size that includes the normalization of the deviation vectors and allows the computation of the Lyapunov exponents spectrum via QR factorization,
    \item Computation of the error in each integration step for the adaptive step size ODE solvers,
    \item Computation of the GALI indicator using Singular Value Decomposition (SVD) \cite{Skokos2008},
    \item Computation of the SALI indicator \cite{Skokos2001,Skokos2003,Skokos2004},
    \item Computation of the chaos indicators based on Lagrangian Descriptors \cite{daquin22,Hille22,zimper23,caliman2025,jiménezlópez2025},
    \item Generation of neighbouring trajectories for the calculation of the chaos indicators based on Lagrangian Descriptors.
\end{itemize}

\section{Examples} \label{sec:examp}

Aiming to test the different functionalities presented in Sec.(\ref{sec:func}) that the package offers, we will simulate the trajectories of three widely known Hamiltonian systems that exhibit a great variety of dynamical behaviours, allowing our method's performance to be tested across different scenarios. 

The three systems that we have chosen for our examples are the dimensionless double pendulum \cite{jimenez2024}, the H\'enon-Heiles system \cite{henon1964applicability} and the Fermi-Pasta-Ulam system \cite{fermi1955}. The first two are $2$ DoF systems and will allow us to exemplify how the time of integration per initial condition is reduced as one increments the number of initial conditions that are simultaneously  integrated until the saturation of the GPU is reached; and the Fermi-Past-Ulam system, which has $N$ DoF, will be used to test the behaviour of the functionalities when a high number of DoF is considered.

The Hamiltonian of the dimensionless double pendulum in matrix form has been found to be:

\begin{equation} \label{eq:Ham_DP}
\mathcal{H}(\boldsymbol{\theta},\mathbf{p}) = \mathcal{T}(\boldsymbol{\theta},\mathbf{p}) + \mathcal{V}\left(\boldsymbol{\theta}\right) = \dfrac{1}{2} \mathbf{p}^T B^{-1}(\cos \Delta \theta) \, \mathbf{p} - \alpha(1+\sigma) \cos \theta_1 - \cos \theta_2 \,,
\end{equation}
where $\Delta \theta = \theta_1 - \theta_2$, $\sigma$ is the mass ratio, $\alpha$ is the length ratio and:

\begin{equation}
     B^{-1}(x) = \dfrac{1}{1+\sigma - x^2}\begin{bmatrix}
	\dfrac{1}{\alpha^2} & -\dfrac{x}{\alpha} \\[.35cm]
	-\dfrac{x}{\alpha} & 1+\sigma
\end{bmatrix} \;.
\end{equation}

Then, Hamilton's equations can be written in matrix form as:

\begin{equation} \label{eq:ham_eqs_DP}
\begin{cases}
    \boldsymbol{\theta}^\prime = B^{-1}(\cos \Delta \theta) \, \mathbf{p} \\ 
    \mathbf{p}^\prime = \dfrac{\sin \Delta \theta}{2} \mathbf{p}^T C(\cos \Delta \theta) \, \mathbf{p} \begin{bmatrix}
		1 \\
		-1
	\end{bmatrix} - \begin{bmatrix}
	\alpha(1+\sigma) \sin\theta_1 \\[.1cm]
	\sin\theta_2
	\end{bmatrix}
\end{cases} \, ,
\end{equation}
with $C(x) = -B^{-1} \dfrac{dB}{dx} B^{-1}$.

\begin{figure}
    \centering
    \includegraphics[scale = 0.4]{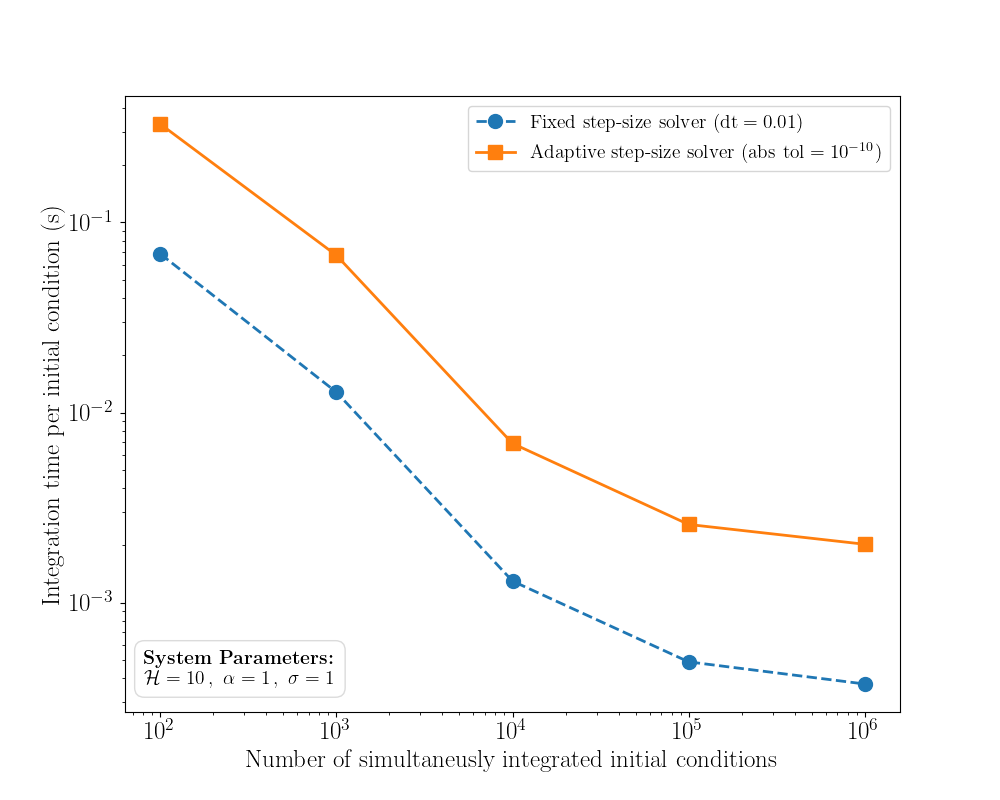}
    \caption{Integration time per initial condition as a function of the number of simultaneously integrated initial conditions for the the dimensionless double pendulum ODE given in Eq.\eqref{eq:ham_eqs_DP} for an integration time $\tau = 10^{3}$. In blue, the results obtained with the fixed step-size solver and in orange the results obtained using the adaptive-size solver. Energy was conserved in all the cases with a precision of $\Delta \mathcal{H} \sim 10^{-9}$.}
    \label{fig:DP}
\end{figure}

In Fig.(\ref{fig:DP}) we have depicted the integration time as a function of the number of simultaneously integrated initial conditions to show the time reduction that is achieved when the local memory of the GPU is used at its peak performance. The comparison between adaptive and fixed step integrators is also shown aiming to highlight the differences between them. While the fixed step solver is faster, the adaptive one provides a more robust approach to analyse systems that might exhibit highly non-linear behaviour. As the package implements both, it is left for the user to decide which one will better fit the case under analysis.

The H\'enon-Heiles system Hamiltonian is:

\begin{equation}  \label{Ham_HH}
    \mathcal{H}(x,y,p_x,p_y) = \dfrac{p_x^2}{2} + \dfrac{p_y^2}{2} +\dfrac{1}{2}(x^2+y^2) + x^2y - \dfrac{1}{3}y^3 \, ,
\end{equation}
and Hamilton's equations of motion for this system are:

\begin{equation} \label{eq:ham_eqs_HH}
\begin{cases}
\dot{x} = \dfrac{\partial \mathcal{H}}{\partial p_x} = p_x \\[.4cm]
\dot{y} = \dfrac{\partial \mathcal{H}}{\partial p_y} = p_y \\[.4cm]
\dot{p}_x = -\dfrac{\partial \mathcal{H}}{\partial x} = - x - 2xy \\[.4cm]
\dot{p}_y = -\dfrac{\partial \mathcal{H}}{\partial y} = - y - x^2 + y^2 
\end{cases}
\;.
\end{equation}

\begin{figure}
    \centering
    \includegraphics[scale = 0.4]{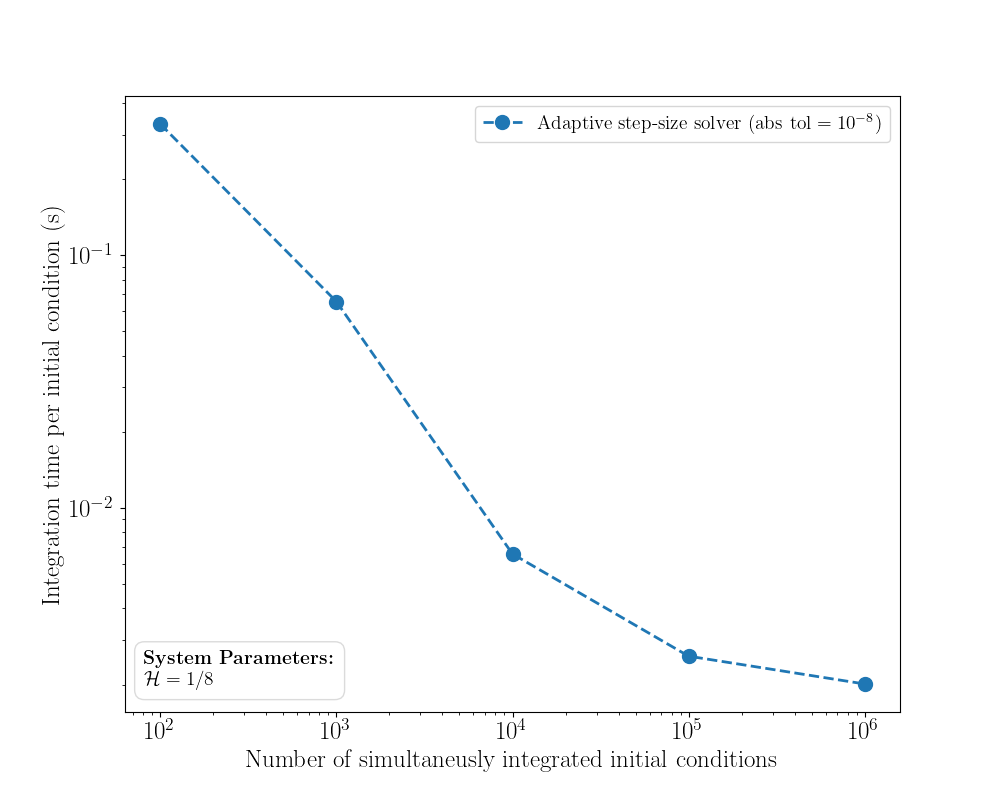}
    \caption{Integration time per initial condition as a function of the number of simultaneously integrated initial conditions for the H\'enon-Heiles system given by Eq.\eqref{eq:ham_eqs_HH} and the evolution of $2$ deviation vectors to calculate the SALI indicator for an integration time $\tau = 10^{4}$ with the adaptive-size solver. Energy was conserved in all the cases with a precision of $\Delta \mathcal{H} \sim 10^{-8}$.}
    \label{fig:HH}
\end{figure}

The H\'enon-Heiles system has been used to analyse how the integration time evolves as a function of the number of simultaneously integrated initial conditions while calculating the SALI indicator. As it has been presented for the Double pendulum in Fig.(\ref{fig:DP}), the results that can be seen in Fig.(\ref{fig:HH}) show the same trend for the H\'enon-Heiles system: a considerable time reduction is observed when the number of simultaneously integrated initial conditions reaches the maximum capacity of the GPU.

Lastly, the Fermi-Pasta-Ulam system is given by:

\begin{equation} \label{eq:ham_FPU}
    \mathcal{H}(\boldsymbol{q}, \boldsymbol{p}, k, \alpha, \beta) = \sum_{i = 1}^{N} \dfrac{p^2_{i}}{2} + \dfrac{k}{2} \sum_{i = 1}^{N} \left( q_{i+1} - q_{i} \right)^2 + \dfrac{\alpha}{3} \sum_{i = 1}^{N} \left( q_{i+1} - q_{i} \right)^3 + \dfrac{\beta}{4} \sum_{i = 1}^{N} \left( q_{i+1} - q_{i} \right)^4
\end{equation}
and Hamilton's equations can be written as:

\begin{equation} \label{eq:ham_eqs_FPU}
    \left\{
    \begin{aligned}
        \dot{q}_i &= \frac{\partial \mathcal{H}}{\partial p_i} = p_i \\[.2cm]
        \dot{p}_i &= - \frac{\partial \mathcal{H}}{\partial q_i} = - k \left[ \Delta q_i - \Delta q_{i+1} \right] - \alpha \left[ (\Delta q_i)^2 - (\Delta q_{i+1})^2 \right] - \beta \left[ (\Delta q_i)^3 - (\Delta q_{i+1})^3 \right]
    \end{aligned}
    \right.
\end{equation}
where $ \Delta q_i = q_i - q_{i-1}$ and $ \Delta q_{i+1} = q_{i+1} - q_i$. For our simulations, we have setted the following boundary conditions over the generalized coordinates:

\begin{equation}
    q_0 = q_{N+1} = 0 \, .
\end{equation}

\begin{figure}
    \centering
    \includegraphics[scale = 0.4]{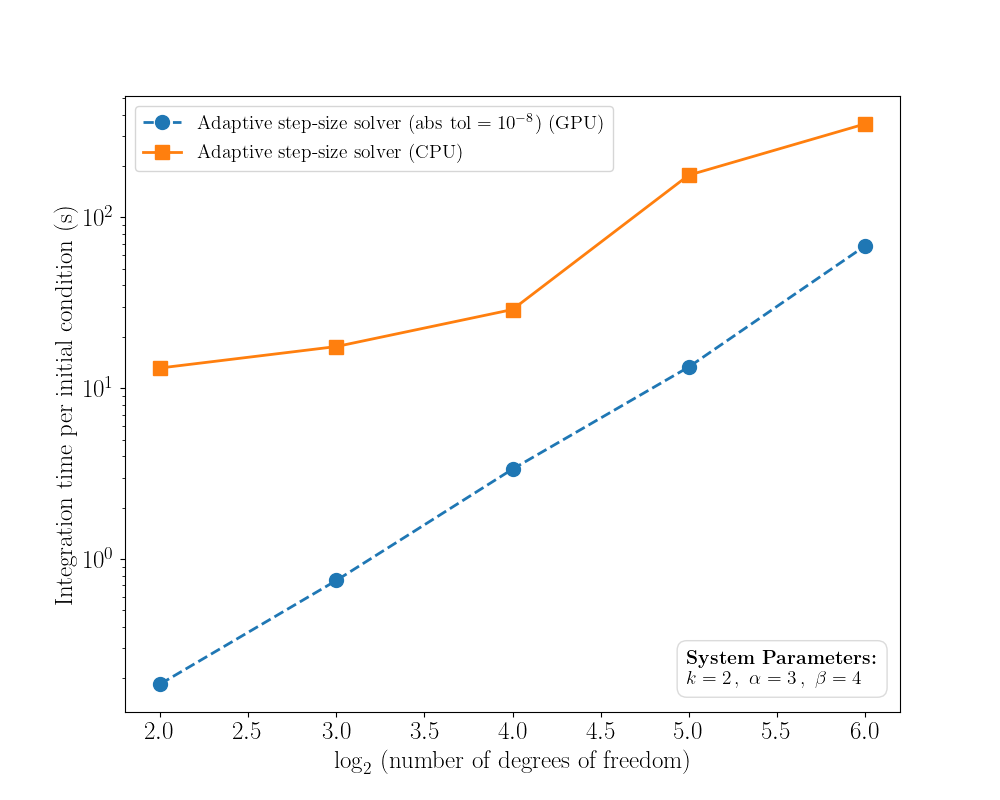}
    \caption{Integration time as a function of the amount of DoF considered for the FPU system given in Eq.\eqref{eq:ham_eqs_FPU} for $10^{3}$ simultaneously integrated randomly generated initial conditions plus the calculation of $\text{GALI}_4$ for an integration time $\tau = 10^3$ time units. In orange, the results obtained performing the integration with the DOP853 solver implemented by \cite{2020SciPy-NMeth} and in blue our implementation in GPU. The absolute tolerance was setted to be $10^{-8}$, providing in all cases a conservation of energy of, at least, $\Delta \mathcal{H} \sim 10^{-8}$.}
    \label{fig:FPU_sim}
\end{figure}

In Fig.(\ref{fig:FPU_sim}) we show the time of integration as a function of the Degrees of Freedom considered for the FPU system for a fixed number of simultaneously integrated initial conditions of $10^{3}$. In it, the comparison between the calculation of GALI$_4$ using CPU and GPU solvers is presented. While for the GPU it exhibits a liner trend, for CPU it seems to be unpredictable. Additionally, our python package also implements the calculation of the Lyapunov exponent spectrum following a very similar approach to the GALI$_k$ computation but using QR factorization instead of the SVD decomposition.

\section{Related research} \label{sec:rr}

The analysis of the underlaying dynamics of ODE systems, specially those that present chaotic regimes, is of great scientific and practical interest across multiple research areas as accelerator and beam dynamics \cite{papaphilippou2014,hwang2019,pocher2024,bartosik2022,montanari2022,li2022,skoufaris2022,montanari2025,wolski2014}, generation of Machine Learning models that can reveal dynamical behaviours \cite{jimenez-lopez2025,greydanus2019,han2021,mattheakis2022,david2023,chen2020,tong2021,tapley2024}, the parametric analysis of physical systems \cite{munoz2007bifurcation,jimenez2024}, astrodynamics \cite{macau2006,colagrossi2022,wakker2015,vallado2001,bate2020,Daquin2019,Daquin2016}, chemical reactions \cite{katsanikas2020,naik2019,galen2017,junginger2016}, Quantum Electrodynamics \cite{Bastarrachea2017}, semiconductors \cite{Eleuch2012}, quantum systems \cite{Macek2011}, nuclear physics \cite{Dietz2017}, lattice dynamics \cite{Sinha1995}, electrodynamical systems \cite{Amarot2025} or galactic dynamics \cite{Zotos2017}. This is just a grasp of the wide range of fields where our package can offer significant improvements to the current state of the art techniques.


\section*{Code availability}

The library presented in this paper and some test cases can be found in \href{https://github.com/JaviJimenezL/Chaoticus}{\textcolor{blue}{Chaoticus}}.


\section*{CRediT authorship contribution statement}

\textbf{Javier Jim\'enez-L\'opez:} Conceptualization, Data curation, Formal analysis, Funding acquisition, Investigation, Resources, Software, Validation, Visualization, Writing - original draft, Writing - review \& editing. \textbf{Jos\'e Bienvenido S\'aez-Landete}: Conceptualization, Data Curation, Funding acquisition, Investigation, Resources, Software, Validation, Visualization. \textbf{V\'{i}ctor J. Garc\'{i}a-Garrido:} Conceptualization, Data curation, Formal analysis, Funding acquisition, Investigation, Project administration, Methodology, Resources, Software, Supervision, Validation, Visualization, Writing - original draft, Writing - review \& editing.

\bibliography{referencias}

\end{document}